\documentclass[12pt,prd,tightenlines,nofootinbib,superscriptaddress,nopacs,nokeys,setspace]{revtex4}
\usepackage[english]{babel}  
\usepackage{xcolor}
\usepackage{bm}
\usepackage{graphics}
\usepackage{rotating}
\usepackage{epsfig}
\usepackage{amsmath}
\usepackage{amsfonts}
\usepackage{setspace}
\usepackage{amssymb}
\usepackage{slashed}
\newcommand{\be}{\begin{equation}}
\newcommand{\ee}{\end{equation}}
\newcommand{\mathsym}[1]{{}}
\newcommand{\unicode}[1]{{}}
\newcommand{\calA}{{\cal A}}
\usepackage{multirow}

\usepackage[unicode=true,hidelinks,colorlinks,linkcolor=black,citecolor={green!70!black},urlcolor={blue!70!black}]{hyperref}

\begin{document}
\normalsize

\title{QCD one-loop correction to Higgs boson decay into quarkonium-pair}
\author{\firstname{I.~N.}~\surname{Belov}}
\email{ilia.belov@cern.ch}
\affiliation{INFN, Sezione di Genova, Italy}

\author{\firstname{A.~V.}~\surname{Berezhnoy}}
\affiliation{SINP MSU, Moscow, Russia}

\author{\firstname{E.~A.}~\surname{Leshchenko}}
\affiliation{Physics department of MSU, Moscow, Russia}

\author{\firstname{A.~K.}~\surname{Likhoded}}
\affiliation{NRC ''Kurchatov Institute'' IHEP, Protvino, Russia}

\begin{abstract}
\small
Rare decays of the Higgs boson into quarkonia pairs are studied within the framework of NRQCD approach. The main decay mechanisms and their interference are studied in detail. One-loop corrections to the widths of these decays are taken into account for the first time.
\end{abstract}


\maketitle

\section{Introduction}
The discovery of the Higgs boson at LHC~\cite{ATLAS:2012yve,CMS:2012qbp} was a triumph for the Standard Model and ushered in a new era in high energy physics, marked by intensive research on the properties of this particle at LHC (see, for example,~\cite{CMS:2022dwd}).
However, for a more detailed investigation of the Higgs boson new facilities are needed. Currently, several projects are discussed, such as ILC, FCC and the muon collider. The energy range announced for FCC is $\sqrt{s}~=~90\div400~\text{GeV}$~\cite{Koratzinos:2014cla} and $\sqrt{s}~=~250~\text{GeV}$ is proposed for ILC~\cite{KEKInternationalWorkingGroup:2019spu}.
In the project of muon collider, it is planned to implement the $\mu^+\mu^-$ collisions at energies from 3~TeV to 14~TeV~\cite{Long:2020wfp}. At all these facilities a detailed study of numerous Higgs boson decays is expected. In this paper we theoretically consider Higgs boson decays into pairs of vector quarkonia within QCD one-loop accuracy. The predicted partial widths of $H\to VV'$ decays are quite small, however, such processes attract the researchers due to their good signatures and high attainable mass resolutions.

\begin{figure}[t]
\centering
\includegraphics[width=\linewidth]{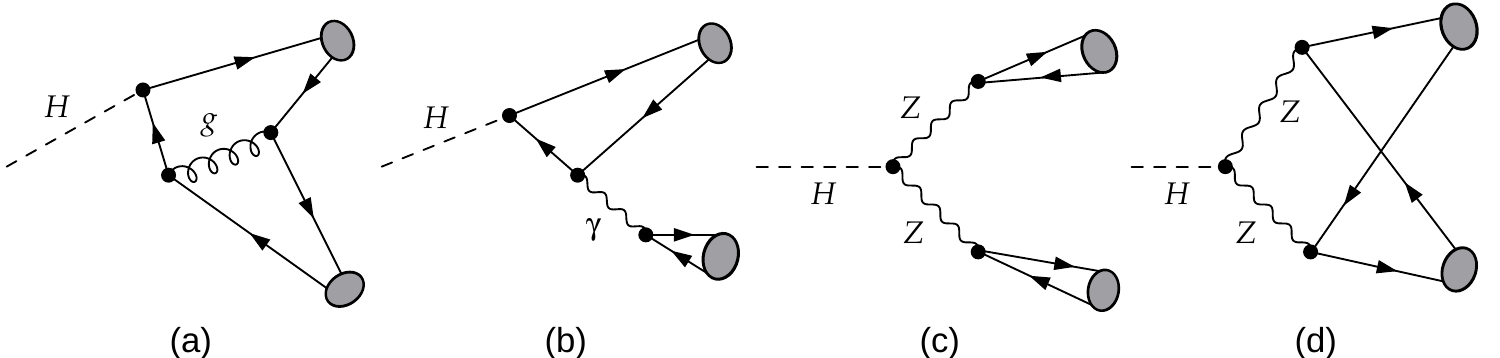}
\caption{Tree-level diagrams for $H\to V V$ decay. 
Several contributions are shown: 
(a)~--~quark-gluon contribution; 
(b)~--~quark-photon contribution; 
(c)~and~(d)~--~$Z$-boson contribution. 
The complete set includes 4 nonzero diagrams of each type (a) and (b) plus 1 nonzero diagram of each type (c) and (d). 
The gray ovals express the quarkonium final states.}
\label{fig:tree}
\vspace{6ex}
\includegraphics[width=\linewidth]{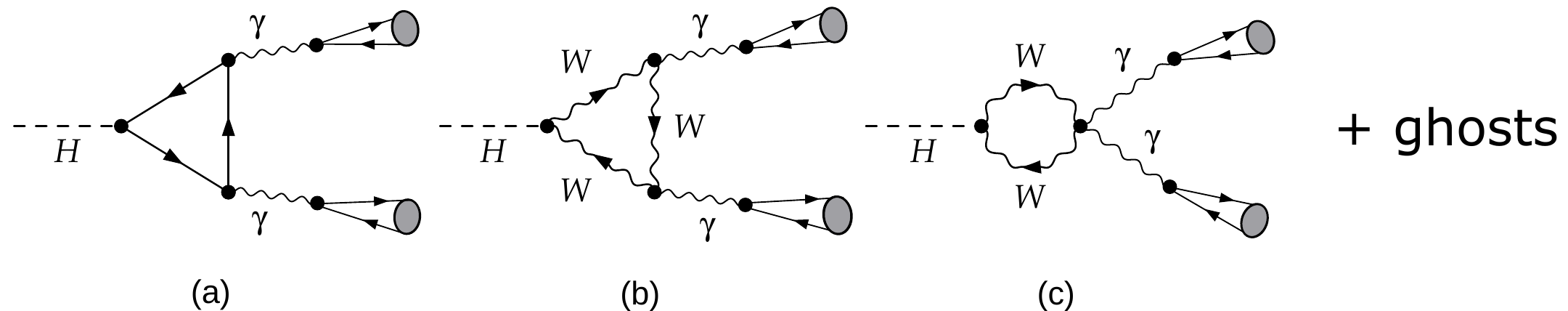}
\caption{Diagrams for $H\to V V$ decay through the subdecay $H\to \gamma^*\gamma^*$. Diagrams include: (a)~---~fermionic loops and (b),\,(c)~--~bosonic loops. Account for the diagrams with ghost loops is a subject of gauge choice. The gray ovals express the quarkonium final states.}
\label{fig:indirect_with_ghosts}
\end{figure}

First estimations of decay width of Higgs boson into a pair of heavy quarkonia were done in~\cite{Bander:1978br} and~\cite{Keung:1983ac}. In study~\cite{Keung:1983ac} the process of Higgs boson decay into quarkonium-pair through the direct interaction of Higgs and a heavy quark was discussed at the lowest QCD order as shown in Fig.\ref{fig:tree}~(a). The study~\cite{Bander:1978br} was devoted to decay into quarkonium-pair in the process $H\to \gamma^* \gamma^*\to V V'$, where the intermediate virtual photons are produced through the fermionic loop as shown in Fig.\ref{fig:indirect_with_ghosts}~(a). These studies have been recently continued in the works~\cite{Kartvelishvili:2008tz,Gao:2022iam,Faustov:2022jfk}.

From theoretical side Higgs boson decay modes with a heavy quarkonium in a final state might be pretty interesting for testing the charm-quark and bottom-quark Yukawa couplings. In this regard two more decay modes must be mentioned: $H\to V\gamma$ and $H\to~V\, Q\bar Q$. Predictions for the $H\to V\,\gamma$ decay have been known for a long time~\cite{Vysotsky:1980cz} and have been recently improved within several approaches~\cite{Bodwin:2013gca,Zhou:2016sot,Brambilla:2019}. The authors of~\cite{Bodwin:2013gca} point out that such decay rates are sensitive to $g_{H Q Q}$ coupling and can be possible for probing this coupling directly at LHC. Nevertheless in such decays the ``indirect'' contribution (which is free of $g_{H Q Q}$ interaction) dominates, and therefore $g_{H Q Q}$ estimation is accessible only through the direct -- indirect interference term. The perspectives for $g_{H Q Q}$ measurement at LHC in the $H\to V\, Q\bar Q$ decay mode are discussed in~\cite{Han:2022rwq}. In the latter case the heavy quark fragmentation decay mechanism obviously dominates. Meantime the presence of two hadronic jets in the inclusive sample makes an experimental analysis rather challenging. Finally we have to stress that neither of mentioned decay modes has been observed so far.

Returning to the decays into quarkonium-pair, one should note that they have a very good experimental signature: $H\longrightarrow V( \to l^+l^-) \; V'(\to l^+l^-)$. For this reason, despite the expected theoretical suppression, these processes are already being sought experimentally. To date, the following upper limits have been achieved by the CMS collaboration \cite{CMS:2019wch,CMS:2022fsq}:
\begin{align*}
\text{Br}(H\to J/\Psi\,J/\Psi) &< 3.8\cdot 10^{-4},\\
\text{Br}(H\to \Upsilon\,\Upsilon) &< 1.7\cdot 10^{-3}.
\end{align*}
Obviously, these searches should be complemented  by the accurate theoretical evaluation within SM. The certain theoretical interest lies in inspecting the example, where the subprocesses, different in terms of the orders by coupling constants, yield comparable contributions to the total amplitude.

In the present work we study the different mechanisms of Higgs decay into the pairs $J/\psi\,J/\psi$, $\Upsilon\,\Upsilon$ and $J/\psi\,\Upsilon$ within the QCD one-loop accuracy and estimate the role of the direct mechanism therein. In the main text we consider the analytic formulas for two identical mesons in a final state while the formulas for $J/\psi\,\Upsilon$ case are given in Appendix.

\section{NRQCD factorization and workflow}

The production of double heavy bound states is effectively described by the NRQCD factorization~\cite{Bodwin:1994jh,Cho:1995ce,Cho:1995vh}. The factorization formalism is introduced to factor out the perturbative degrees of freedom in order to separate the production mechanism into hard (short-distance) and soft (long-distance) subprocesses. Given the fact that $m_Q >> m_Q v$, where $v$ is the velocity of heavy quark in quarkonium, the short-distance interaction corresponds to the perturbative part of $Q\bar Q$-pair production, whereas the long-distance interaction describes the bound state formation and dynamics. 

In our computations of the decay matrix elements we start from the matrix element $H\to Q(p_Q) \bar Q(p_{\bar Q}) Q'(p_{Q'}) \bar Q'(p_{\bar Q'})$ with heavy quarks and antiquarks defined on their mass shells: $p_Q^2=p_{\bar Q}^2=m_Q^2$ and $p_{Q'}^2=p_{\bar Q'}^2=m_{Q'}^2$. As we assign $v=0$ before the projection onto the bound states $V$ and $V'$, each quark carries away half the momentum of the corresponding meson.
To construct the bound states, we replace the spinor products $v(p_{\bar Q})\bar u(p_Q)$ and $v(p_{\bar Q'})\bar u(p_{Q'})$ by the appropriate covariant projectors for color-singlet spin-triplet states with a zero relative velocity as per
\begin{align}
\label{proj}
     &\Pi_{V}(P,m)=\frac{\slashed P- m}{2\sqrt{2}}\ \slashed \epsilon \otimes \frac{\boldsymbol 1}{\sqrt{N_c}}\,, 
     &\Pi_{V'}(P',m')=\frac{\slashed P'- m'}{2\sqrt{2}}\slashed \epsilon' \otimes \frac{\boldsymbol 1}{\sqrt{N_c}}\,, 
 \end{align}
where $P$ and $P'$ are momenta of the mesons, $m=2m_Q$, $N_c=3$, $\varepsilon$ and $\varepsilon'$ are polarizations of the vector mesons, satisfying the following constraints: $\epsilon\cdot \epsilon^* =-1$, $\epsilon\cdot P=0$, $\epsilon'\cdot \epsilon'^* =-1$, $\epsilon'\cdot P'=0$.

The factorized matrix element within the framework of NRQCD has the form specified as below 
\begin{equation}
{\cal A}_{VV'} = \frac{\langle {\cal O}_{V}\rangle^{1/2} \langle {\cal O}_{V'}\rangle^{1/2}}{\sqrt{m m'} \;N_c}  {\cal M}^{\mu\nu}\left(P,P'\right)\epsilon_{\mu}\epsilon_{\nu}'
\label{eq:ampO}\,,
\end{equation}
where ${\cal M}^{\mu\nu}\left(P,P'\right)\epsilon_{\mu}\epsilon_{\nu}'$
is the perturbative matrix element of Higgs decay into the two quark-antiquark pairs, projected onto the two vector quark-antiquark states with momenta $P$ and $P'$ by means of the projectors~\eqref{proj}; $\langle {\cal O}_{V}\rangle$ and $\langle {\cal O}_{V'}\rangle$ are vacuum-saturated analogs of the NRQCD matrix element $\langle  O(^3S_1)\rangle$ defined in~\cite{Bodwin:1994jh}. Using the relation to the wave function at origin $\Psi(0)^2=  \frac{1}{2N_c}\langle{\cal O}\rangle$ one can rewrite~(\ref{eq:ampO}) through $\Psi_V(0)$, $\Psi_{V'}(0)$ as follows:
\begin{equation}
{\cal A}_{VV'} = \Psi_{V}(0) \, \Psi_{V'}(0) \; \sqrt{\frac{4}{m m'}}
\;{\cal M}^{\mu\nu}\left(P,P'\right)\epsilon_{\mu}\epsilon_{\nu}'\label{eq:ampPsi}. 
\end{equation}
For compatibility with the previous articles on $H\to VV$ decay study~\cite{Keung:1983ac,Bander:1978br,Kartvelishvili:2008tz,Faustov:2022jfk,Gao:2022iam} in the text we follow the form of writing~\eqref{eq:ampPsi}.

The computation of tree-level diagrams is organized with the help of two packages: \texttt{FeynArts}~\cite{Hahn:2000kx} for generation of matrix elements and \texttt{FeynCalc}~\cite{Shtabovenko:2020gxv} for further symbolic calculation. For the computation of one-loop diagrams we use a more complicated toolchain: \texttt{FeynArts}~\cite{Hahn:2000kx} $\to$ \texttt{FeynCalc}~\cite{Shtabovenko:2020gxv} $\to$ \texttt{FIRE}~\cite{Smirnov:2008iw} $\to$ \texttt{X}~\cite{Patel:2016fam}.

To calculate the amplitudes with a loop  we set the relative momenta of heavy quark-antiquark motion inside quarkonia to zero before carrying out loop integrations rather than resorting to much more expensive matching calculations (see \cite{Beneke:1997zp} for details). After the projection onto the bound states, traces calculation and calculation of color factors the tensor integrals are reduced to scalar ones by means of Passarino-Veltman reduction. The infrared divergent integrals are optionally simplified by partial decomposition of fractions to integrals with fewer propagators. The \texttt{FIRE} package provides the complete reduction of the obtained integrals to master integrals, using the IBP reduction strategy based on the Laporta algorithm~\cite{Laporta:2000dsw}. After the \texttt{FIRE} reduction only one-, two- and three-point master integrals $\boldsymbol{A}_0$, $\boldsymbol{B}_0$, $\boldsymbol{C}_0$ are left in the amplitudes. The master integrals are handled in the representation of \texttt{X} package.

\section{Contributions without a gluon loop}
\label{sec:LO}
There are two tree-level contributions to the process under study, described by the diagrams, in which a pair of heavy quarks is being produced in a direct interaction with a Higgs boson. In case of $H\to V \gamma$ or $H\to V\, Q\bar Q$ decays these contributions are known from literature as ``direct'' ones~\cite{Bodwin:2013gca,Zhou:2016sot,Brambilla:2019,Han:2022rwq}. In the first contribution a pair of heavy quarks emits a virtual gluon: $H\to Q \bar Q +g (\to [Q' \bar Q']_\textrm{\bf{8}})$ as shown in Fig.~\ref{fig:tree}(a). In this study we will refer to this mechanism as a quark-gluon one. In the second contribution a pair of heavy quarks emits a virtual photon: $H\to Q \bar Q +\gamma (\to [Q' \bar Q']_\textrm{\bf{1}})$  as shown in Fig.~\ref{fig:tree}(b). We will refer to the last mechanism as a quark-photon one. Despite the fact that the amplitudes of these two contributions are proportional to different coupling constants ($\alpha_s$ and $\alpha$ respectively), their values are comparable due to the different structures of propagators.

The contribution of quark-gluon and quark-photon subprocesses to the Higgs boson decay width into two equivalent mesons $V$ has a simple analytic form~\cite{Keung:1983ac,Gao:2022iam}:
\begin{equation}
\label{eq:direct}
\Gamma_{VV}^\textrm{dir} = \frac{128 \sqrt{2} \pi  G_F |\Psi_V(0)|^4 \sqrt{r^2/4-1}\, \left(r^4 - 4r^2 + 6\right)}{9\, r^7 M_H^3}\bigg[8 \alpha_s  +9 \alpha e_q^2 r^2 \bigg]^2,
\end{equation}
where $r = \frac{M_H}{m_V}=\frac{M_H}{2m_Q}$. The small value of $\alpha$ is compensated by the large coefficient $r^2$: $r^2 \sim 10^3$ for charmonium-pair and $r^2 \sim 10^2$ for bottomonium-pair. As seen from~(\ref{eq:direct}), the amplitudes of two regarded mechanisms interfere constructively.

According to our estimations, the quark-gluon contribution is essential only in case of decay into $\Upsilon\, \Upsilon$. Obviously, the quark-gluon mechanism is absent in case of the decay into $J/\psi\, \Upsilon$ --- a pair of quarkonia with different hidden flavours. The quark-photon mechanism essentially contributes to both decays into $J/\psi\, J/\psi$ and $\Upsilon\, \Upsilon$ and dominates in the decay into $J/\psi\, \Upsilon$. The estimated decay widths for each of the discussed contributions are presented in Table~\ref{tab:higgs_nlo_res}.

Except for the direct decay mechanisms there are two decay mechanisms going through the intermediate Higgs decay into two virtual bosons: either photons or $Z$-bosons. With respect to the direct ones these two decay mechanisms might be called as ``indirect''. Let us move to their detailed consideration.

First indirect decay mechanism considers Higgs decay into two virtual photons via a fermionic or bosonic loop: $H\to \gamma^* \gamma^*\to V V'$, as shown in Fig.~(\ref{fig:indirect_with_ghosts}). We will refer to this mechanism as an EW-loop one. In case of the $H\to \gamma \gamma$ decay into two real photons the loop factor $J(r_f^2, r_W^2)$ is well known~\cite{ellis:1976,Shifman:1979eb,Spira:1995rr}. Under approximation of negligible virtuality of intermediate photons we have derived the following expression for Higgs decay width into two equivalent mesons $V$:
\begin{equation}
\label{eq:indirect}
\Gamma^\textrm{EW-loop}_{VV} = \frac{18\sqrt{2}\, G_F\, \alpha^4 e_Q^4\, |\Psi_V(0)|^4 \sqrt{r^2/4-1}}{\pi M_H^3}\, r\left(r^4-4r^2+6\right) \big|J(r_f^2, r_W^2)\big|^2,
\end{equation}
where $J(r_f^2, r_W^2)\sim 1$ is a complex valued loop factor accounted for quark and $W$-boson loops: $J(r_f^2, r_W^2)=\sum\limits_{f} N_c e_f^2 A_f(r_f^2) + A_W(r_W^2)$ and $r_f=\frac{M_H}{2m_f}$, $r_W=\frac{M_H}{2M_W}$. 
See, for example,~\cite{Spira:1995rr} and formulas (1--4) therein for definition of $A_f$ and $ A_W$ functions.

Since formulas (\ref{eq:direct}) and (\ref{eq:indirect}) have been derived from amplitudes with the same Lorentz structures, they can be easily joined as per
\begin{multline}
\label{eq:dir_indir}
\Gamma^\textrm{dir + EW-loop}_{VV} = \frac{128 \sqrt{2} \pi  G_F |\Psi_V(0)|^4 \sqrt{r^2/4-1}\, \left(r^4 - 4r^2 + 6\right)}{9\, r^7 M_H^3}\times \\
\times\bigg|8 \alpha_s  +9 \alpha e_Q^2 r^2 + \frac{9}{8\pi}\alpha^2 e_Q^2 r^4\, J(r_f^2,r_W^2)\bigg|^2.
\end{multline}
Equation~(\ref{eq:dir_indir}) demonstrates that the EW-loop contribution is enhanced by a factor $\alpha^2 r^4$ with respect to the quark-photon one. Bearing in mind $\alpha r^2 \sim 10$ for charmonium-pair and $\alpha r^2 \sim 1$ for bottomonium-pair one might conclude that in case of the decay into $J/\psi\, J/\psi$ this contribution is to be dominant. Our numerical estimations confirm this suggestion (see Table~\ref{tab:higgs_nlo_res}).

It should be noted that the naive consideration of loop contributions described above leads to the loss of additional structure in the amplitude, which appears due to the virtuality of intermediate photons. Within this approach the amplitude corresponding to the decay width~(\ref{eq:indirect}) can be expressed as per
\begin{equation}
\label{eq:H2realphotons_str}
{\cal A}^\textrm{EW-loop}_{VV} \sim  J(r_f^2,r_W^2)\left\{\epsilon^* \cdot {\epsilon'}^* \;\frac{M_H^2-2m_V^2}{2 m_V^2}-\frac{P_V\cdot{\epsilon'}^*}{m_V} \;\frac{P_{V'}\cdot\epsilon^*}{m_V}\right\}.
\end{equation}
In contrast, the rigorous consideration of virtual photons gives rise to a more complex amplitude, which consists of two structures:
\begin{multline}
\label{eq:H2virtualphotons_str}
{\cal A}^\textrm{EW-loop}_{VV} \sim  \widetilde{J}(r_f^2,r_W^2,r^2)\left\{\epsilon^* \cdot {\epsilon'}^* \; \frac{M_H^2-2m_V^2}{2 m_V^2}-\frac{P_V\cdot{\epsilon'}^*}{m_V} \; \frac{P_{V'}\cdot\epsilon^*}{m_V}\right\}
+\\+
K(r_f^2,r_W^2,r^2)\left\{\frac{P_{V} \cdot {\epsilon'}^* }{m_V}-\frac{ P_{V'}\cdot {\epsilon}^* }{m_V} \; \frac{M_H^2-2m_V^2}{2m_V^2}\right\} \left\{\frac{P_{V'} \cdot {\epsilon}^* }{m_V}-\frac{P_{V}\cdot {\epsilon'}^* }{m_V} \; \frac{M_H^2-2m_V^2}{2m_V^2}\right\}.
\end{multline}
The loop factors $\widetilde{J}(r_f^2,r_W^2,r^2)$ and $K(r_f^2,r_W^2,r^2)$ have been obtained within the already mentioned  toolchain \texttt{FeynArts} $\to$ \texttt{FeynCalc} $\to$ \texttt{FIRE} $\to$ \texttt{X}.

As it is shown in \cite{Bodwin:2013gca} and as confirmed by our calculations, the structure~(\ref{eq:H2realphotons_str}) is enough to describe the amplitude of the process $H\to V \gamma$ with an accuracy up to ${\cal O}(1/r^2)$. However that is wrong for the process $H\to VV'$, where both the structures~(\ref{eq:H2virtualphotons_str}) should be taken into account. According to our exact calculation, accounting for the second structure, proportional to $K(r_f^2,r_W^2,r^2)$, increases the loop contribution approximately by 20\%. The analogous comment has been made in the paper~\cite{Faustov:2022jfk}, where the discussed processes were studied within the relativistic quark model. We point out that in Table~\ref{tab:higgs_nlo_res} we provide the correct values using the complete expression~(\ref{eq:H2virtualphotons_str}) for EW-loop amplitude.

The second indirect decay mechanism for the studied process originates from Higgs decay into two virtual $Z$-bosons: $H\to Z^*Z^* \to VV'$, as shown in Fig.~\ref{fig:tree}(c,d). We will refer to this mechanism as a vector boson one. Note, that unlike the EW-loop decay mechanism, where essential is only the diagrams with each of the final mesons being formed in a $\gamma^* \to V$ transition (see Fig.~\ref{fig:indirect_with_ghosts}), a contribution of the diagram (d) in Fig.~\ref{fig:tree} is not suppressed, since the propagators of intermediate bosons are massive ($M_Z \sim M_H$, compare two terms in parentheses in~(\ref{eq:h2VV_zz})).

The expression for decay width into two equivalent mesons $V$ through the vector boson mechanism can be written out as follows:
\begin{multline}
\label{eq:h2VV_zz}
    \Gamma_{VV}^\textrm{vb} = \frac{16\sqrt{2}\pi G_F \alpha^2 |\Psi_V(0)|^4 \sqrt{r^2/4-1}\, M_Z^4 M_H\left(r^4/4-r^2+3\right)}{r^3\,   \sin^4 2\theta_W}\times \\
    \times\bigg[\frac{6 (g_v^Q)^2}{(M_Z^2-m_V^2)^2} + \frac{(g_v^Q)^2 + (g_a^Q)^2}{(M_Z^2-M_H^2/4)^2} \bigg]^2,
\end{multline}
where $g_v^Q = T_3^Q-2e_Q \sin^2\theta_W$ and $g_a^Q = T_3^Q$, $e_Q$ -- the quark charge, $T_3^Q$ -- the third component of weak isospin and $\sum_{\text{spin}}|\epsilon^*\cdot\epsilon^{'*}|^2 = \left(r^4/4 - r^2 + 3\right)$. Such a contribution dominates in case of decay into $\Upsilon\,\Upsilon$, while in case of decay into $J/\psi\,J/\psi$ or $J/\psi\,\Upsilon$ it appears to be negligible with respect to the above-mentioned contributions.

It is meaningful to mention the signs of the interference terms between all the discussed contributions. The total decay width is proportional to
\begin{equation}\label{eq:interference}
\Gamma \sim  \big|\calA^{Q\,g}+ \calA^{Q
\, \gamma} +\calA^\textrm{vb} + \calA^\textrm{EW-loop}\big|^2.
\end{equation}
In the right hand side of~(\ref{eq:interference}) the amplitudes $\calA^\textrm{vb}$ and $\calA^\textrm{EW-loop}$ (as well as $\calA^{Q\,g}$ and $\calA^{Q\,\gamma}$) interfere constructively with each other. The rest four direct -- indirect interference terms are destructive: $2\,\textrm{Re}\left[\calA^{Q\,g}\cdot \calA^\textrm{EW-loop} \right] < 0$, etc. A more particular conclusion about negative interference between EW-loop and quark-photon amplitudes has been made in~\cite{Bodwin:2013gca,Brambilla:2019} with regard to $H\to V\gamma$ decay.

\section{Gluon loop corrections}
\label{sec:NLO}
QCD corrections to the EW-loop decay mechanism are fairly well studied within the Higgs decay into two photons~(see for example the first works~\cite{Djouadi:1991,Dawson:1993}). At present the loop factor $J(r_f^2,r_W^2)$ entering~\eqref{eq:H2realphotons_str}, is known through three loops by $\alpha_s$~\cite{Davies:2021}. The correction size is well under control and restricted to $\,< 2\%$. 

\begin{figure}[t]
\centering
\begin{minipage}[h]{\linewidth}
\includegraphics[width=\linewidth]{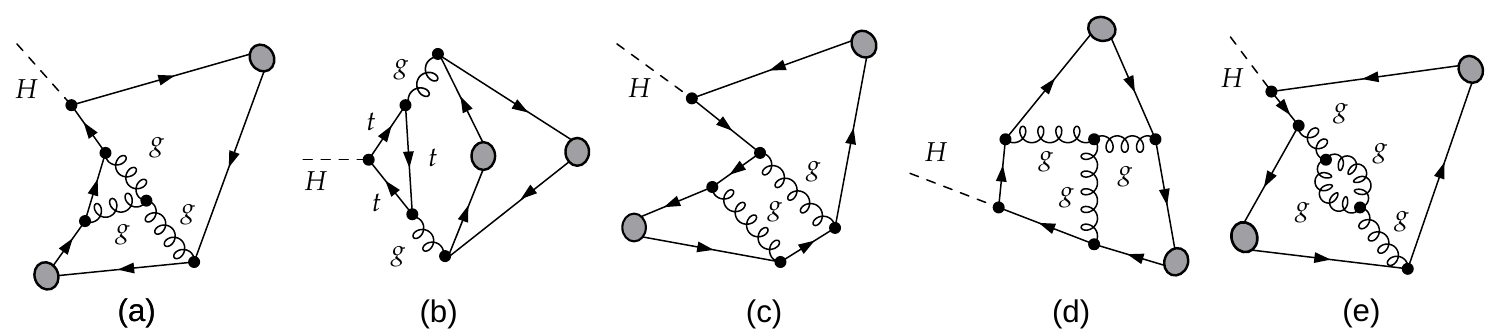}
\caption{Five sample diagrams for quark-gluon mechanism at the next-to-leading order by $\alpha_s$.
The complete set includes 80 nonzero diagrams.}
\label{fig:higgs_qcd_nlo}
\end{minipage} 
\begin{minipage}[h]{\linewidth}
\vspace{4ex}
\includegraphics[width=\linewidth]{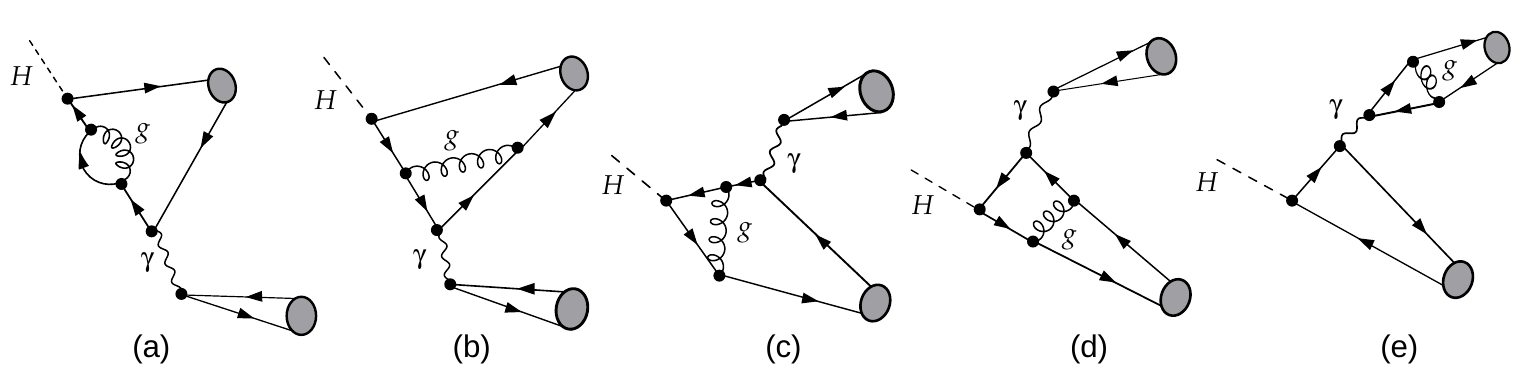}
\caption{Five sample diagrams for quark-photon mechanism at the next-to-leading order by $\alpha_s$.
The complete set includes 20 nonzero diagrams.}
\label{fig:higgs_qed_nlo}
\end{minipage}
\begin{minipage}[h]{\linewidth}
\vspace{5ex}
\includegraphics[width=\linewidth]{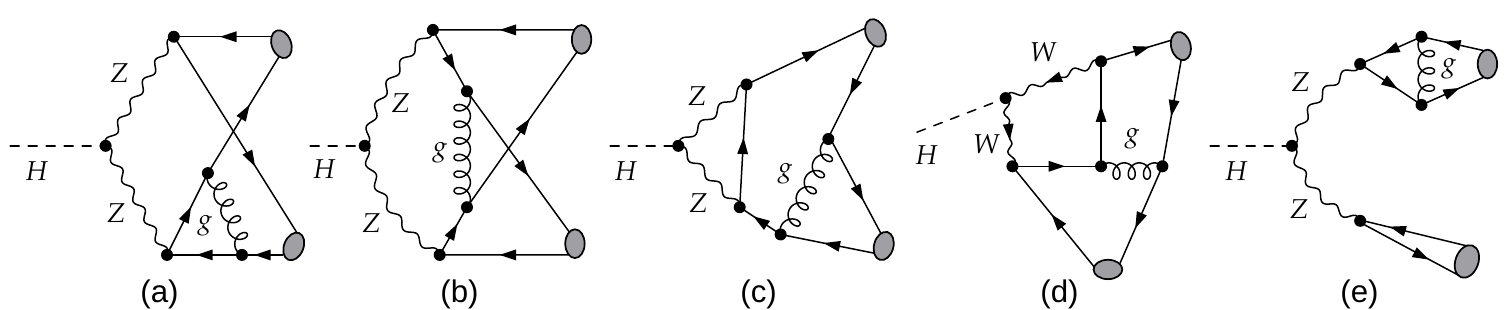}
\caption{Five sample diagrams for vector boson mechanism at the next-to-leading order by $\alpha_s$.
The complete set includes 20 nonzero diagrams.
}
\label{fig:higgs_zz_nlo}
\end{minipage}
\end{figure}

QCD corrections to the direct decay mechanisms are known only for decay $H\to V\gamma$ driven by the quark-photon mechanism~\cite{Vysotsky:1980cz,Zhou:2016sot}. In this study we concentrate on the estimation of one-loop QCD corrections to the quark-gluon, quark-photon and vector boson mechanisms contributing to $H\to VV'$ decay. The examples of the diagrams describing these corrections are schematically shown in Figs~\ref{fig:higgs_qcd_nlo}, \ref{fig:higgs_qed_nlo}, \ref{fig:higgs_zz_nlo}.\,\footnote{Note, that in our study we don't single out a contribution of a top quark to the decay like it was considered, for example, in~\cite{Kartvelishvili:2008tz}. Instead we include the diagram~(b) in Fig.~\ref{fig:higgs_qcd_nlo} into ${\cal A}_{NLO}^{Q\,g}$ as a one-loop correction to the quark-gluon mechanism.}

At the next-to-leading order we evaluate three sets of diagrams, whose amplitudes are designated as $\calA_\textrm{NLO}^{Q\, g}$,  $\calA_\textrm{NLO}^{Q\, \gamma}$, $\calA_\textrm{NLO}^\textrm{vb}$. In order to distinguish the one-loop corrections to the tree-level amplitudes from amplitude of the EW-loop mechanism in the text, we stick to the term ``gluon loop'' correction. Hence amplitudes for all the contributions, shown in Figs~\ref{fig:tree}--\ref{fig:higgs_zz_nlo}, are combined as
\begin{align}
&{\cal A}=\calA_{LO}^{Q\,g}+ \calA_{LO}^{Q
\, \gamma} +\calA_{LO}^\textrm{vb} + \calA^\textrm{EW-loop}, \\
&\calA_\textrm{\,gluon loop}= \calA_{NLO}^{Q\, g}+\calA_{NLO}^{Q\, \gamma}+ \calA_{NLO}^\textrm{vb},
\end{align}
where subscripts ``LO'' and ``NLO'' refer to the $\alpha_s$ expansion.

Within the quark-photon decay mechanism the decay $H\to V\,\gamma^*$ is followed by the transition $\gamma^*\to V$. These two subprocesses are entirely separated and obviously take place at different scales, so it seems reasonable to treat the gluon corrections to the hard part of the amplitude and to the relatively soft part of the amplitude separately. We take into account the ${\cal O}(\alpha_s)$ correction only to the hard part of the amplitude, i.e. to the decay $H\to V\,\gamma^*$. The diagrams of type (e) in Fig.~\ref{fig:higgs_qed_nlo} have been skipped in our estimations, insofar as such diagrams can be interpreted as a correction to the wave function of the final quarkonium. Nevertheless it is not entirely clear whether a similar factorization takes place against the $Z^*\to V$ subprocess in the vector boson mechanism. We don't skip the diagrams of type~(e) in Fig.~\ref{fig:higgs_zz_nlo} in our estimations given the presence of massive propagator with $M_Z \sim M_H$. However, in the latter case the question of including the ${\cal O}(\alpha_s)$ correction into the quarkonium wave function requires further consideration.

In the studied decay the one-loop QCD corrections are entirely manifested in the interference term between real and virtual amplitudes:
\begin{equation}
\label{eq:A_tot}
    |\calA^\textrm{tot}|^2 = 
    |\calA|^2
    + 2\,\textrm{Re}
    \left[\calA\cdot \calA_\textrm{\,gluon loop}\right].
\end{equation}
The correction for soft gluon radiation is not the case since the final states are in color singlets. The finite expression for ${\cal A}_\textrm{gluon loop}$ is obtained by the application of renormalization procedure.

The next-to-leading order calculation technique was reported in our previous paper~\cite{Berezhnoy:2021tqb}, devoted to ${\cal O}(\alpha_s)$ correction to quark-gluon mechanism in quarkonium-pair production. Here we briefly repeat the renormalization technique.

In semi-automatic computations the calculation of ${\cal A}_{NLO}$ starts with the generation of the NLO matrix elements ${\cal \widetilde A}_{NLO}$ with physical values of masses and spinors. Algebraic transformations and evaluation of Feynman integrals are carried out in the conventional dimensional regularization (CDR) scheme where all the momenta (loop and external) as well as Dirac matrices are defined in $D=4-2\varepsilon$ dimensional space.\,\footnote{It is worth to note  that $\calA_{NLO}^\textrm{vb}$ is the only term to which a trace with a single $\gamma_5$ matrix  contributes generating  a structure proportional to $\varepsilon^{p_V \; p_{V'} \; \epsilon \; \epsilon'}$.  This structure disappears after summation over polarizations in \eqref{eq:A_tot}. For this reason, it is safe to discard  the traces with a single $\gamma_5$, and thus avoid the problem  with definition  of $\gamma_5$ in $D$ dimensions.} Within the dimensional regularization singular and regular parts of the amplitudes are isolated. The divergences are further canceled with counter-terms built from the leading-order amplitudes so that ${\cal A}_{NLO}= {\cal \widetilde A}_{NLO}+{\cal A}_{CT}$ yields a finite expression for the renormalized amplitude.

The renormalization procedure implies that masses and spinors are renormalized in the so-called on-shell scheme and the coupling constant is renormalized in $\overline{MS}$ scheme:
\begin{equation}
\begin{array}{lcl}
\displaystyle
\label{eq:renorm_const}
 Z_{m}^{OS} &=&\displaystyle 1 - 3C_F\,\frac{\alpha_s}{4\pi}\left[\frac{1}{\epsilon}  - \gamma_E + \ln\left(\frac{16\pi\mu_R^2}{m^2}\right) + \frac{4}{3}\right] + {\cal O}(\alpha_s^2), \\
 \\
 Z_{2}^{OS} &=&\displaystyle 1 - 3C_F\,\frac{\alpha_s}{4\pi} \left[\frac{1}{\epsilon} - \gamma_E + \ln\left(\frac{16\pi\mu_R^2}{m^2}\right) + \frac{4}{3}\right] + {\cal O}(\alpha_s^2), \\
 \\
Z_g^{\overline{MS}} &=&\displaystyle 1 - \frac{\beta_0}{2}\frac{\alpha_s}{4\pi}\left[\frac{1}{\epsilon} -\gamma_E + \ln(4\pi)\right] + {\cal O}(\alpha_s^2),
\end{array}
\end{equation}
where $m=m_V$; $\epsilon = \epsilon_{UV} = \epsilon_{IR}$~(infrared and ultraviolet divergences are evaluated within a single code); $C_F = \frac{N_c^2-1}{2N_c} = 4/3$, $\beta_0 = \frac{11}{3}N_c -\frac{2}{3}N_f$, $\gamma_E$ is the Euler constant and $\mu_R$ is a renormalization scale.

The divergent parts of the next-to-leading order amplitudes carry only the poles $1/\epsilon$. The counter-terms for their subtraction are constructed according to:

\begin{equation}
\label{eq:renorm}
\begin{array}{lcl}
 \delta Z_2^2 \calA_{LO}^{Q\, g} \Biggr|_{\substack{\boldsymbol{ m \to \delta Z_m m} \\\boldsymbol{g_s \to \delta Z_g g_s}}} &= &{\cal A}_{CT}^{Q\, g} + {\cal O}(\alpha_s^3) + \ldots, \\
 \\
\delta Z_2\,{\cal A}_{LO}^{Q\,\gamma} \Biggr|_{\substack{\boldsymbol{ m \to\: \delta Z_m\,m}}} &= & {\cal A}_{CT}^{Q\,\gamma} + {\cal O}(\alpha_s^2) + \ldots, \\
\\
\delta Z_2^2\,{\cal A}_{LO}^\textrm{vb} \Biggr|_{\substack{\boldsymbol{ m \to\: \delta Z_m\,m}}} & = &{\cal A}_{CT}^\textrm{vb} + {\cal O}(\alpha_s^2) + \ldots,
\end{array}
\end{equation}
where in the right hand side ${\cal A}_{CT}^{Q\, g}$ is the lowest order term ${\cal O}(\alpha_s^2)$ in expansion by $\alpha_s$, whereas ${\cal A}_{CT}^{Q\,\gamma}$ and ${\cal A}_{CT}^\textrm{vb}$ are the lowest order terms ${\cal O}(\alpha_s)$ in expansion by $\alpha_s$.

\section{$K$-factors}
In order to study the gluon loop corrections in detail, we calculate the $K$-factors for the quark-gluon, quark-photon and vector boson decay mechanisms separately. The obtained $K$-factors can be represented as per
\begin{equation}
\label{eq:kfac}
\begin{array}{lcccl}
K^{Q\, g} &=& 1+2\,\textrm{Re}\big[\calA_{LO}^{Q \, g}\cdot \calA_{NLO}^{Q\, g}\big] \big/ |\calA_{LO}^{Q \, g}|^2 &=& 1 + \alpha_s(\mu)\left(c_1+ c_0\ln(\mu/M_H)\right),\\
\\
K^{Q\, \gamma} &=& 1+2\,\textrm{Re}\big[\calA_{LO}^{Q\, \gamma}\cdot \calA_{NLO}^{Q\, \gamma}\big] \big/ |\calA_{LO}^{Q \, \gamma}|^2&=& 1 - c_2\,\alpha_s(\mu),\\
\\
K^\textrm{vb}&=& 1+2\,\textrm{Re}\big[\calA_{LO}^\textrm{vb}\cdot \calA_{NLO}^\textrm{vb}\big] \big/ |\calA_{LO}^\textrm{vb}|^2 &=& 1 + c_3\,\alpha_s(\mu),
\end{array}
\end{equation}
where $c_0>0$, $c_1>0$, $c_2>0$; $c_3>0$ for charmonium-pair final state and $c_3<0$ for bottomonium-pair final state. The $K$-factors for the quark-photon and vector boson mechanisms are free of logarithmic terms $\sim\ln(\mu)$ originating from Feynman integrals, as such terms cancel out under renormalization. Thus, $K^{Q\, \gamma}$ and $K^\textrm{vb}$ depend on scale only through the strong coupling $\alpha_s(\mu)$.

In~\eqref{eq:kfac} the coefficient $c_0$ has the simplest form: $c_0 = \beta_0/\pi$, whereas the expressions for $c_1$, $c_2$ and $c_3$ are too cumbersome to present their complete analytical form in the text. Nevertheless we write out the approximate expressions for coefficients $c_1$, $c_2$ under expansion over the small parameters.

Neglecting the masses of light quarks and keeping the masses of $c$-, $b$-, $t$-quarks in the loops, we expand $c_1$ in a series by $1/r^2 = m_V^2/M_H^2$ and $1/r_{b,c}^2 = (2m_{b,c})^2/M_H^2$. The following expression has been obtained for the leading term:
\begin{multline}\label{eq:c1-exp}
    c_1 = \frac{1}{18 \pi}\Bigg[
    \frac{36}{r_t^5}\left(1 - r_t^2\right)^{3/2}\arctan\left(\frac{2r_t\sqrt{1-r_t^2}}{1-2r_t^2}\right)
    +\\+ \frac{6}{r_t^5}\left(r_t^4+8r_t^2-6\right)\left(4-r_t^2\right)^{1/2}\arctan\left(\frac{r_t\sqrt{4-r_t^2}}{2-r_t^2}\right) 
    - \frac{6}{r_t^2} - 12\ln\left(r_t\right)
    +\\+ 36 \ln ^2(r) - 168\ln(2)\ln(r) + 84\ln(r) -\\- 84\ln^2(2) + 366\ln(2) - 11\pi^2 - 39 \Bigg]
    + {\cal O}(1/r_b^2,1/r_c^2,1/r^2) + \ldots
\end{multline}
In~(\ref{eq:c1-exp}) our standard notation $r_t = M_H/(2m_t)$ is used for a top quark, which contribution is clearly seen.

The leading term in the expansion of $c_2$ with respect to $1/r^2$ reads
\begin{equation}\label{eq:c2-exp}
c_2 = \frac{16 \ln(2)}{3\pi} \ln(r)+\frac{4\pi ^2 + 42 + 36\ln^2(2) -24\ln(2)}{9 \pi } + {\cal O}(1/r^2) + \ldots
\end{equation}
The leading logarithmic term $16\ln(2)\ln(r)/(3\pi)$ in the expansion~\eqref{eq:c2-exp} is the same as that one obtained in~\cite{Vysotsky:1980cz} for ${\cal O}(\alpha_s)$ correction to the direct decay $H\to V\,\gamma$. This serves as an additional cross-check for our calculations.

\begin{figure}[t]
\centering
\includegraphics[width=1\linewidth]{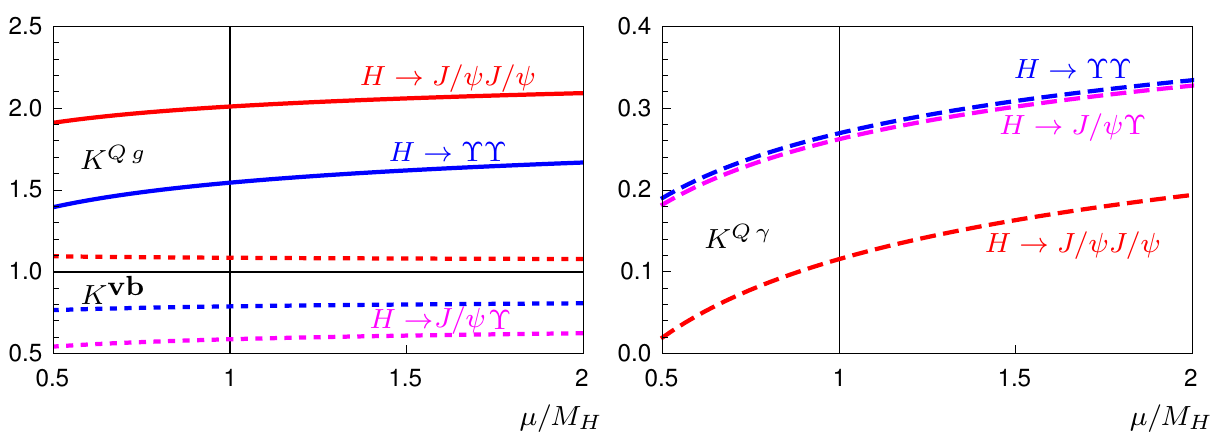}
\caption{The scale dependence of $K$-factors. Left panel: quark-gluon mechanism (solid curves) and vector boson mechanism (dashed-dotted curves). Right panel: quark-photon mechanism (dashed curves).} 
\label{fig:kfactorsVV}
\end{figure}

The coefficient $c_3$ does not acquire a compact approximate form under expansion over~$1/r$, and therefore we present $K^\textrm{vb}$ graphically in Fig.~\ref{fig:kfactorsVV} along with $K^{Q\,g}$ and $K^{Q\,\gamma}$.

For numerical estimations we make use of the strong coupling constant within a two-loop accuracy and a number of flavors $N_f = 6$. The reference value is $\alpha_s(M_Z) = 0.1129$. Considering the quark-gluon mechanism, we use the same scale $\mu$ both for renormalization scale and the strong coupling scale, which is reflected in~\eqref{eq:kfac}.

The gluon loop correction to the quark-gluon mechanism is large and positive both for decays into $J/\psi\,J/\psi$ and $\Upsilon\,\Upsilon$. As seen from Fig.~\ref{fig:kfactorsVV}~(left), the factor $K^{Q\,g}$  changes in the ranges $1.9 \div 2$\,~($J/\psi\, J/\psi$~mode) and  $1.3\div 1.6$\,~($\Upsilon\,\Upsilon$~mode) with respect to variation of the scale $\mu$ from $M_H/2$ to $2 M_H$. These values are consistent with ones obtained for the $H\to B_c^*\:B_c^*$ decay reported in our previous work~\cite{Belov:2021toy}.

The gluon loop correction to the quark-photon mechanism is large and negative for all the considered decay modes (see Fig.~\ref{fig:kfactorsVV}~(right)). 
For the decays into $\Upsilon \Upsilon$-pair and into $J/\psi\,\Upsilon$-pair the values of $K^{Q\,\gamma}$ change in the range $0.18 \div 0.33$ with respect to scale variation from $M_H/2$ to $2M_H$. For the decay into $J/\psi\, J/\psi$ this $K$-factor changes from $0.02$ at $\mu = M_H/2$ to $0.19$ at $\mu = 2M_H$. This circumstance suggests that the one-loop accuracy might not be sufficient to describe this contribution. Approximately the same values for the analogous $K$-factor have been obtained in work~\cite{Zhou:2016sot}, where the quark-photon mechanism is considered for the direct decay $H\to J/\psi\,\gamma\: \left(\Upsilon\,\gamma\right)$. The results agreement serves as a good check since in both studies the diagrams of the same topology are regarded at the next-to-leading order.

The gluon loop correction to the vector boson mechanism behaves differently for each of the investigated processes (see Fig.~\ref{fig:kfactorsVV}~(left)). It is small and positive for the decay into $J/\psi\, J/\psi$. In this case the correction size is restricted to $10$\% within the mentioned scale range. For two remaining decay modes $\Upsilon\, \Upsilon$ and $J/\psi\,\Upsilon$ the correction appears to be negative and considerable. The corresponding $K^\textrm{vb}$ values lie in the ranges $0.76 \div 0.81$ and $0.54 \div 0.62$.

\section{Decay widths} 
\label{sec:app}
We present the numerical results for decay widths in Table~\ref{tab:higgs_nlo_res}\,.\,\footnote{The width values and appropriate values of branching fractions are calculated using the Standard Model expectation of the total Higgs boson width $\Gamma_H=4.1$~MeV and recommended reference value of Higgs boson mass $M_H=125.1$~GeV~\cite{CERN:2017}.} By the main contributions we mean the contributions, introduced in Section~\ref{sec:LO}. By the corrections we mean the interference terms $2\,\textrm{Re}[\calA \cdot \calA_{NLO}^{Q\, g}]$, $2\,\textrm{Re}[\calA \cdot \calA_{NLO}^{Q\, \gamma}]$ and $2\,\textrm{Re}[\calA \cdot \calA_{NLO}^\textrm{vb}]$, where ${\cal A}$ is the total amplitude of the main contribution. As could be expected for a highly suppressed decay into quarkonium-pair, the widths are small. The obtained values are of the order $10^{-11} \div 10^{-12}$ in GeV units.

It is notable that for each decay mode the four considered decay mechanisms contribute differently to the total decay width. In case of $J/\psi\,J/\psi$ mode the EW-loop decay mechanism dominates against all the others. In case of $\Upsilon\,\Upsilon$ mode the vector boson decay mechanism dominates against all the others. In case of $J/\psi\,\Upsilon$ mode the quark-photon and EW-loop mechanisms are dominant. Moreover the quark-gluon mechanism is suppressed (absent) for all the cases.

In Table~\ref{tab:higgs_nlo_res} one can see the size and the sign for each the QCD one-loop correction with respect to including the diagrams with a gluon loop in Figs~\ref{fig:higgs_qcd_nlo},~\ref{fig:higgs_qed_nlo},~\ref{fig:higgs_zz_nlo}. In case of $J/\psi\, J/\psi$ decay mode the ${\cal O}(\alpha_s)$ correction is mainly driven by the constructive interference with NLO quark-photon amplitude ($+23$\%), where the correction size is mostly determined by the interference term $2\,\textrm{Re}[\calA^{\textrm{EW-loop}}\cdot\calA_{NLO}^{Q\, \gamma}]$. In case of $\Upsilon\,\Upsilon$ decay mode the ${\cal O}(\alpha_s)$ correction is mainly driven by the destructive interference with NLO vector boson and NLO quark-photon amplitudes ($-17$\% and $-8.2$\% correspondingly). In this instance the correction size is mostly gained through the interference terms $2\,\textrm{Re}[\calA_{LO}^{\textrm{vb}}\cdot\calA_{NLO}^{Q\,\gamma}]$ and $2\,\textrm{Re}[\calA_{LO}^{\textrm{vb}}\cdot\calA_{NLO}^{\textrm{vb}}]$. Concerning the $J/\psi\,\Upsilon$ decay mode, the ${\cal O}(\alpha_s)$ correction is mainly driven by the destructive interference with NLO quark-photon amplitude ($-49$\%), where the correction size is mostly gained through the interference terms $2\,\textrm{Re}[\calA_{LO}^{Q\,\gamma} \cdot \calA_{NLO}^{Q\, \gamma}]$ and $2\,\textrm{Re}[\calA^{\textrm{EW-loop}} \cdot \calA_{NLO}^{Q\,\gamma}]$.

\begin{table}[t]
\caption{Decay widths at the next-to-leading approximation by $\alpha_s$ in units $10^{-12}$ GeV. The values are calculated at scale $\mu = M_H$ with wave functions~\cite{Eichten:2019hbb} and masses $m_{J/\psi}=3.10~\text{GeV}$, $m_{\Upsilon}=9.46~\text{GeV}$.}
\begin{ruledtabular}
\begin{tabular}{|cll|clclcl|}
\hline
\multicolumn{2}{|c|}{\multirow{2}{*}{mechanism}} &
  \multirow{2}{*}{Figure} &
  \multicolumn{6}{c|}{width in $10^{-12}$ GeV at $\mu=M_H$} \\ \cline{4-9} 
\multicolumn{2}{|c|}{} &
   &
  \multicolumn{2}{l|}{$H\to J/\psi\:J/\psi$} &
  \multicolumn{2}{l|}{$H\to \Upsilon\:\Upsilon$} &
  \multicolumn{2}{l|}{$H\to J/\psi\:\Upsilon$} \\ \hline
\multicolumn{1}{|l|}{\multirow{4}{*}{main contribution}} &
  \multicolumn{1}{l|}{quark-gluon} &
  $\,1$\,(a) &
  $8.34 \cdot 10^{-4}$ &
  \multicolumn{1}{l|}{} &
  $\phantom{-}0.22$ &
  \multicolumn{1}{l|}{} &
  --- &
   \\
\multicolumn{1}{|l|}{} &
  \multicolumn{1}{l|}{quark-photon} &
  $\,1$\,(b) &
  $\phantom{2}2.25$ &
  \multicolumn{1}{l|}{} &
  $\phantom{-}0.42$ &
  \multicolumn{1}{l|}{} &
  $\phantom{-}5.61$ &
   \\
\multicolumn{1}{|l|}{} &
  \multicolumn{1}{l|}{vector boson} &
  $\,1$\,(c,d) &
  $\phantom{2}0.48$ &
  \multicolumn{1}{l|}{} &
  $\phantom{-}3.82$ &
  \multicolumn{1}{l|}{} &
  $\phantom{-}0.19$ &
   \\
\multicolumn{1}{|l|}{} &
  \multicolumn{1}{l|}{EW-loop} &
  $\,2$ &
  $29.04$ &
  \multicolumn{1}{l|}{} &
  $\phantom{-}0.06$ &
  \multicolumn{1}{l|}{} &
  $\phantom{-}3.48$ &
   \\ \hline
\multicolumn{3}{|c|}{\textbf{total}} &
  $\mathbf{19.62}$ &
  \multicolumn{1}{l|}{} &
  $\phantom{-}\mathbf{4.90}$ &
  \multicolumn{1}{l|}{} &
  $\phantom{-}\mathbf{3.02}$ &
   \\ \hline
\multicolumn{1}{|c|}{\multirow{3}{*}{correction}} &
  \multicolumn{1}{l|}{quark-gluon} &
  $\,3$ &
  $-0.64$ &
  \multicolumn{1}{r|}{\footnotesize ($-3.3$\%)} &
  $-0.09$&
  \multicolumn{1}{r|}{\footnotesize ($-1.8$\%)} &
  --- &
  \multicolumn{1}{r|}{\footnotesize ---\phantom{\%}} \\
\multicolumn{1}{|c|}{} &
  \multicolumn{1}{l|}{quark-photon} &
  $\,4$ &
  $\phantom{-}4.57$ &
  \multicolumn{1}{r|}{\footnotesize ($+23$\phantom{.}\%)} &
  $-0.40$ &
  \multicolumn{1}{r|}{\footnotesize ($-8.2$\%)} &
  $-1.48$ &
  \multicolumn{1}{r|}{\footnotesize ($-49\phantom{.}$\%)} \\
\multicolumn{1}{|c|}{} &
  \multicolumn{1}{l|}{vector boson} &
  $\,5$ &
  $\phantom{-}0.16$ &
  \multicolumn{1}{r|}{\footnotesize ($+0.8$\%)} &
  $-0.84$ &
  \multicolumn{1}{r|}{\footnotesize $(-17\phantom{.}$\%)} &
  $-0.27$ &
  \multicolumn{1}{r|}{\footnotesize ($-8.9$\%)} \\ \hline
\multicolumn{3}{|c|}{\textbf{total (with corrections)}} &
  $\mathbf{23.72}$ &
  \multicolumn{1}{l|}{} &
  $\phantom{-}\mathbf{3.57}$ &
  \multicolumn{1}{l|}{} &
  $\phantom{-}\mathbf{1.27}$ &
   \\ \hline
\end{tabular}
\end{ruledtabular}
\label{tab:higgs_nlo_res}
\end{table}

Account for QCD one-loop correction allows one to stabilize the scale dependence in the quark-gluon contribution. The quark-photon and vector boson contributions at the ${\cal O}(\alpha_s)$ accuracy remain sensitive to the scale choice. The EW-loop contribution in our estimations is taken without the QCD correction and thus does not carry the dependence on $\mu$. Combining all the contributions into the total decay width we obtain different in size uncertainties for different decay modes. In addition the widths are very sensitive to the wave functions accuracy as far as proportional to their fourth power. 

The calculated widths under the variation of $\mu$ from $M_H/2$ to $2M_H$ are: 
\begin{equation}
\begin{array}{lcl}
\Gamma\left(J/\psi\,J/\psi\right) &= &\left|\frac{\Psi_{J/\psi}}{\Psi_{J/\psi, 0}}\right|^4\cdot(2.33\div 2.39)\cdot 10^{-11}~\text{GeV}, \\
\Gamma\left(\Upsilon\,\Upsilon\right) &= &\left|\frac{\Psi_{\Upsilon}}{\Psi_{\Upsilon, 0}}\right|^4 \cdot (3.40\div 3.70)\cdot 10^{-12}~\text{GeV},\\
\Gamma\left(J/\psi\,\Upsilon\right) &= &\left|\frac{\Psi_{J/\psi}}{\Psi_{J/\psi, 0}}\right|^2\left|\frac{\Psi_{\Upsilon}}{\Psi_{\Upsilon, 0}}\right|^2\cdot (1.08\div 1.43)\cdot 10^{-12}~\text{GeV},
\end{array}
\label{eq:widths}
\end{equation}
where we have normalized the wave functions to the values from~\cite{Eichten:2019hbb}. In our predictions account for ${\cal O}(\alpha_s)$ corrections increases the width of the decay $H\to J/\psi\,J/\psi$ by $(19\div 22)\%$, decreases the width of decay $H\to \Upsilon\,\Upsilon$ by $(25\div 30)\%$ and decreases the width of the $H\to J/\psi\,\Upsilon$ process by $(50\div 60)\%$.

\section{Summary}
In this paper we have theoretically studied the $H\to J/\Psi\, J/\Psi$, $H\to \Upsilon\, \Upsilon$ and $H\to J/\Psi\, \Upsilon$ decays within the framework of the NRQCD approach. The tree-level contributions of quark-photon, quark-gluon and vector boson mechanisms to these decays have been considered along with the QCD one-loop corrections. The latter have been calculated for the first time. EW-loop contribution and its interference with the other contributions have been  also  taken into account. The accounted QCD one loop corrections increase the width of the decay $H\to J/\psi\,J/\psi$ by $(19\div 22)\%$ and  decrease the widths  of  $H\to \Upsilon\,\Upsilon$ and $H\to J/\psi\,\Upsilon$ decays by $(25\div 30)\%$ and $(50\div 60)\%$, correspondingly. It has been shown that the mechanisms of direct decay, which involve the $g_{HQQ}$ interaction, do not dominate in the considered processes.
 
The following ranges for branching fractions in the Standard Model have been obtained: 
\begin{equation}
\begin{array}{lcl}
\text{Br}\left(J/\psi\:J/\psi\right) &= & (5.82\div 5.98)\cdot 10^{-9}, \\
\text{Br}\left(\Upsilon\:\Upsilon\right) &= & (8.49\div 9.24)\cdot 10^{-10}, \\
\text{Br}\left(J/\psi\:\Upsilon\right)  &= & (2.71\div 3.57)\cdot 10^{-10}.
\end{array}
\label{eq:branchings}
\end{equation}

The results are directly related to the searches of rare $H$-boson decays into double quarkonia states in LHC detectors~\cite{CMS:2019wch,CMS:2022fsq}. However, the predicted width values are too small, and therefore the observation of such decays under the conditions of current experiments will mean the manifestation of physics beyond the Standard Model.

\acknowledgments
The authors are grateful to  A. P. Martynenko and F. A. Martynenko for fruitful discussions.

I.~N.~Belov acknowledges support by the Foundation for the Advancement of Theoretical Physics and Mathematics ``BASIS'' (grant No. 20-2-2-2-1) during his employment in Lomonosov Moscow State University. The work of A.~V.~Berezhnoy and E.~A.~Leshchenko was carried out within the framework of the scientific program of the National Center for Physics and Mathematics, the project “Particle Physics and Cosmology”.

\appendix*
\section{Decay to charmonium-bottomonium pair}
In this appendix we provide the main formulae for $H\to J/\psi\,\Upsilon$ decay, which are not listed in the body text. We denote the two vector mesons as $V$ and $V'$ and use the same mass ratios $r=\frac{M_H}{m_V}$ and $r'=\frac{M_H}{m_{V'}}$ as in the body text.

For the main contributions to decay width (no gluon loop) one reads
\begin{multline}
\label{eq:direct_CB}
    \Gamma_{VV'}^\textrm{dir} = 
    \frac{128 \sqrt{2} \pi G_F \alpha^2 |\Psi_{V}(0)|^2 |\Psi_{V'}(0)|^2 \sqrt{r^2 r'^2-(r+r')^2}}{9 M_H^3 r^2 r'^2}
    \times \\ \times  \left(\frac{r^2 r'^2 \left(r^2+r'^2\right)+\left(r^2-r'^2\right)^2}{r^4r'^4 - \left(r^2 - r'^2\right)^2}\right)^2 \left(\left(r^2+r'^2-r^2r'^2\right)^2+2 r^2 r'^2\right),
\end{multline}
\begin{multline}
\label{eq:indirect_CB}
    \Gamma_{VV'}^\textrm{EW-loop} = 
    \frac{8\sqrt{2} G_F \alpha ^4  |\Psi_{V}(0)|^2 |\Psi_{V'}(0)|^2
    \sqrt{r^2 r'^2-(r+r')^2}\: \big|J(r_f^2, r_W^2)\big|^2 }{9 \pi M_H^3 r^2 r'^2}
    \times \\ \times \left(\left( r^2+ r'^2-r^2r'^2\right)^2+2 r^2 r'^2\right),
\end{multline}
\begin{multline}
\label{eq:h2VV_zz_CB}
    \Gamma_{VV'}^\textrm{vb} = 
    \frac{144 \sqrt{2} \pi \alpha ^2 G_F |\Psi_{V}(0)|^2 |\Psi_{V'}(0)|^2 (g_v^Q)^2 (g_v^{Q'})^2 M_H M_Z^4 \sqrt{r^2 r'^2-(r+r')^2}}
    {(M_Z^2-m_{V}^2)^2 (M_Z^2-m_{V'}^2)^2\: r^4 r'^4 \sin^4 2\theta_W} \times \\ \times \left(\left(r^2+r'^2-r^2 r'^2\right)^2+8 r^2 r'^2\right).
\end{multline}
Note that expressions~\eqref{eq:direct_CB},\eqref{eq:indirect_CB},\eqref{eq:h2VV_zz_CB} are symmetric with respect to $r \longleftrightarrow r'$ permutation, but these expressions do not turn into the analogous expressions~\eqref{eq:direct},\eqref{eq:indirect},\eqref{eq:h2VV_zz} for two identical mesons under $r' = r$ replacement.

The approximate expression for $c_2$ coefficient, comprising the gluon loop correction to the quark-photon mechanism,  takes the form
\begin{equation}
\label{eq:с2_CB}
    c_2' = 
    \frac{16\ln(2)}{3\pi}\left[\frac{ r^2 \ln(r')+ r'^2\ln (r)}{r^2+r'^2}\right]+
    \frac{4\pi^2 + 42 -12 \ln^2(2)-24 \ln(2)}{9 \pi} +  {\cal O}(1/r^2,1/r'^2) + \ldots,
\end{equation}
which differs from the corresponding coefficient~\eqref{eq:c2-exp} for two identical mesons under $r=r'$ replacement.

The coefficient $c_3$, comprising the gluon loop correction to the vector boson mechanism, is trivial in case of $J/\psi\,\Upsilon$ mode: 
\begin{equation}
    c_3' = -\frac{32}{3\pi}.
\end{equation}
It includes only correction caused by the diagrams of type~(e) in Fig.~\ref{fig:higgs_zz_nlo}.

\bibliography{main}
\end{document}